\documentstyle[preprint,aps,epsf]{revtex}
\begin{document} 
\begin{flushright}
DOE/ER/40561-21-INT98\\
LBNL-42172
\end{flushright}
\begin{center}
{\Large \bf Enhancement of Intermediate Mass Dileptons From Charm Decays At SPS
Energies}\\[2ex]
{Ziwei Lin and Xin-Nian Wang}\\[2ex]
{Nuclear Science Division, LBNL, Berkeley, CA 94720\\
and\\
Institute for Nuclear Theory, University of Washington, Box 351550, 
Seattle, WA 98195}\\[2ex]
{\today}
\end{center}

\begin{abstract}
We study the dimuon yield from open charm decays in the intermediate mass
region(IMR) in heavy-ion collisions at the CERN-SPS energy. We find that 
final state rescatterings which broaden the $m_T$ spectrum of charmed 
mesons can enrich the part of the phase space covered by the NA50 experiment,
thus leading to an apparent enhancement of IMR dileptons within the acceptance
of the experiment.  Such an enhancement increases with the dimuon invariant
mass and the single muon energy cut-off. 
\vspace{0.5in}
\end{abstract}


Dileptons are important signals in high-energy heavy-ion collisions. Depending
on their invariant masses they can probe the collision processes at different
stages. Low invariant-mass dileptons are predominantly produced at the late
stage of the collisions. So they provide information about the medium
modification of hadron properties such as mass and spectral
functions\cite{brown,hl}.   
A large enhancement of low invariant-mass dileptons has indeed been observed
in HELIOS-3\cite{helios3} and CERES\cite{ceres} experiments. Many theoretical
studies attributed this enhancement either to a dropping
$\rho$-mass in a medium of high baryon-density\cite{ko}, or to the 
collisional broadening of hadronic spectral functions\cite{spectral}.

At large invariant masses, dileptons are mostly produced from partonic
processes in the early stage of heavy-ion collisions. For example, dileptons
from the Drell-Yan processes\cite{dy} dominate the large invariant mass region
at the CERN-SPS energy. Thus they can provide a benchmark for parton luminosity
or the centrality of the collisions in which proposed signals of the phase
transition to the quark-gluon plasma such as $J/\psi$ suppression\cite{jpsi}
and thermal 
dileptons\cite{sx} can be studied via the dilepton channel. However, at
collider energies such as the BNL RHIC and CERN LHC, the Drell-Yan processes no
longer dominate dilepton productions at reasonably large invariant
masses. Dileptons from decays of charm and bottom mesons instead dominate
\cite{vogt}.  These large invariant-mass dileptons from heavy quark
decays may also give us information about energy loss of heavy quarks traveling
through the dense medium\cite{shuryak,loss,kampfer}.

As one can naively expect, both partonic and hadronic processes will contribute
to dileptons with invariant masses between the above noted regions, i.e., in
the intermediate mass region (IMR). 
An enhancement of intermediate-mass dileptons
was observed in $S+W$ collisions at HELIOS-3\cite{helios3} and was attributed
to secondary 
meson-meson interactions by Gale and Li \cite{li-gale}. NA38/NA50 experiments
\cite{carlos,na-imr} observed similar enhancement in the mass region $1.5$
GeV/c$^2 < M_{\mu^+\mu^-} < 2.5$ GeV/c$^2$. The dimuon invariant-mass spectrum
in $p+p$ collisions is fitted by a combination of dileptons from Drell-Yan and
decays of charmed mesons. Assuming a scaling of binary collisions for hard
processes, it was found that dimuon spectra in $p+A$ collisions can be
well accounted for but results from $S+U$ and especially $Pb+Pb$ collisions
indicate significant enhancement \cite{na-imr,na50int}. It is difficult to
argue that the excess of IMR dileptons comes from enhanced charm production
according to previous estimates \cite{pre,thermal} in which it is concluded
that the charm production from secondary parton scattering is negligible with a
realistic estimate of temperature and gluon fugacity at both RHIC and LHC
energies.  Even if a quark-gluon plasma could be formed at the CERN-SPS energy,
the temperature and gluon fugacity are probably too low to produce any
significant amount of charmed quarks. The charm production in a hadronic gas
can also be neglected. 

We focus in this paper on another possible scenario in which IMR dileptons can
be enhanced in the acceptance of NA50 $Pb+Pb$ experiment. We demonstrate that
the modification of the phase space distribution of D-mesons due to final state
rescatterings can change the shape of the dilepton distribution in phase
space. Since most experiments only measure a small part of the whole phase
space, the redistribution of the dileptons from decays of charmed mesons can
give rise to an apparent enhancement of IMR dileptons within the limited
acceptance of the experiments.  


In the dense medium formed during heavy-ion collisions, interactions among
hadrons or partons will lead to  partial thermalization of the system. Thus,
effects of strong final state rescatterings are expected, e.g., in the
experimentally measured particle spectra in transverse momentum
\cite{tflow}. These spectra can be parameterized in an exponential form,
$dN/m_T d m_T \propto e^{-(m_T-m)/T_{eff}}$, and are characterized by the
inverse slope $T_{eff}$. It is found that $T_{eff}$ increases with the system 
size\cite{nu,na49qm97} and is roughly a linear function of the particle mass.
For example, $T_{eff}$ for protons increases from $148$ MeV in
$pp$ collisions to $208$ MeV in central $SS$ collisions, and to $289$ MeV in
central $Pb+Pb$ collisions at measured by NA44\cite{na44prl}. 
The increase is often
attributed to a collective transverse flow, which essentially comes from
final state rescatterings among secondary particles\cite{ollitrault}.
Such effect of final state rescatterings which causes the broadening of
hadronic $m_T$ spectra, could also happen to $D$-mesons($D^0,\bar
{D^0},D^+,D^-$ etc).

To obtain a qualitative feature of this effect, we model final state
rescatterings in the following way.  We assume that D-mesons thermalize
with their {\em local} environment, where the degree of thermalization is
parameterized by a temperature $T$.  Due to the lack of knowledge
about interactions between D-mesons and other hadrons, one cannot accurately
estimate the value of $T$. We assume that D-meson spectrum is influenced
by final state rescatterings in the same way as other hadrons. Thus,
$T_{eff}$ for D-mesons in central $Pb+Pb$ collisions is enhanced to about the
same value as those for protons and $\phi$-mesons ($290$ MeV).   
From this $T_{eff}$, the parameter $T$ can be constrained to be about $150$ MeV
in our model. Note that $T$ is the local temperature which differs from the
definition of $T_{eff}$. Later in the discussion we will show the dependence of
our results on $T$.


We calculate the open charm production in $pp$ collisions with the latest
PYTHIA program\cite{pythia}.  We use the MRS D$-^\prime$\cite{mrsd-p} parton
distribution functions with intrinsic parton transverse momentum $k_t$ of
$1$ GeV/c, and with $m_c=1.3$ GeV/c$^2$. The charm pair cross section in $pp$
collisions, $\sigma^{pp}_{c\bar c}$, is normalized to $9.5\pm2 \mu$b at
$E_{Lab}=200$ GeV\cite{carlos}, which then gives $6.8\mu$b at $E_{Lab}=158$
GeV.  For central $Pb+Pb$ collisions at $158$ AGeV, the number of charm pairs
is then given by  
\begin{equation}
N^{PbPb}_{c\bar c}=\sigma^{pp}_{c\bar c}\;\; T_{PbPb}(0), \nonumber
\end{equation}
where the overlap function $T_{PbPb}(0)$ is taken to be
$30.4/$mb\cite{tab}.  

We take the Peterson fragmentation function to describe the hadronization of
charmed quarks to D-mesons:
\begin{equation}
D(z) \propto \frac {1}{z [1-1/z-\epsilon/(1-z)]^2} \nonumber 
\end{equation}
where $z=p_D/p_c$, and $\epsilon=0.06$.
D-mesons are then assumed to be thermalized in its local frame with a
temperature parameter $T$ as the following.  
For a D-meson with an initial four-momentum $p_{D,\alpha}=(E_D, \vec
p_D)$ in the Lab frame, first a local random momentum in the D-meson rest frame
is generated from the thermal distribution $dN/d^3p^l_D \propto e^{-E^l_D/T}$.
Then the local four-momentum $(E^l_D, \vec {p^l_D})$ is boosted back to the Lab
frame, i.e., to the frame 
with $\gamma_D=1/\sqrt {1-\beta^2_D}$, where $\vec \beta_D=\vec p_D/E_D$.
This will simulate a locally thermal D-meson spectrum with some collective
flow in both transverse and longitudinal directions. The local temperature $T$
is taken to be $150$ MeV if not specified otherwise in this paper.
  
\begin{figure}[h]
\setlength{\epsfxsize=\textwidth}
\setlength{\epsfysize=0.6\textheight}
\centerline{\epsffile{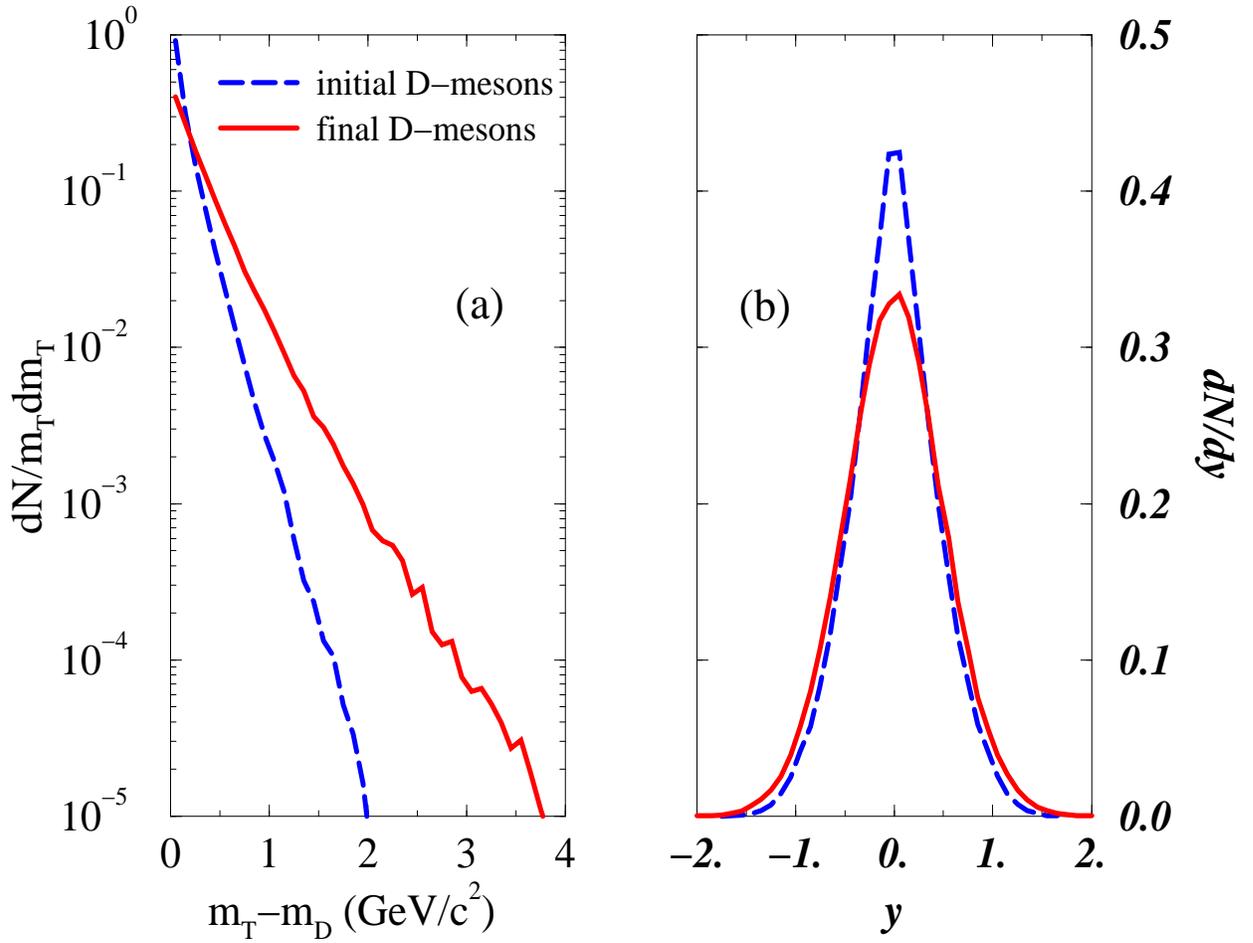}}
\vspace{1cm}
\caption{
(a) $m_T$ spectra of initial and final D-mesons at mid-rapidity.
(b) Rapidity spectra of initial and final D-mesons in the CMS frame.
The dashed curve represents initial D-mesons after the Peterson fragmentation,
and the solid curve represents final D-mesons after the thermalization with
$T=150$ MeV.
}
\label{fig1_mty}
\end{figure}

\begin{figure}[h]
\setlength{\epsfxsize=\textwidth}
\setlength{\epsfysize=0.6\textheight}
\centerline{\epsffile{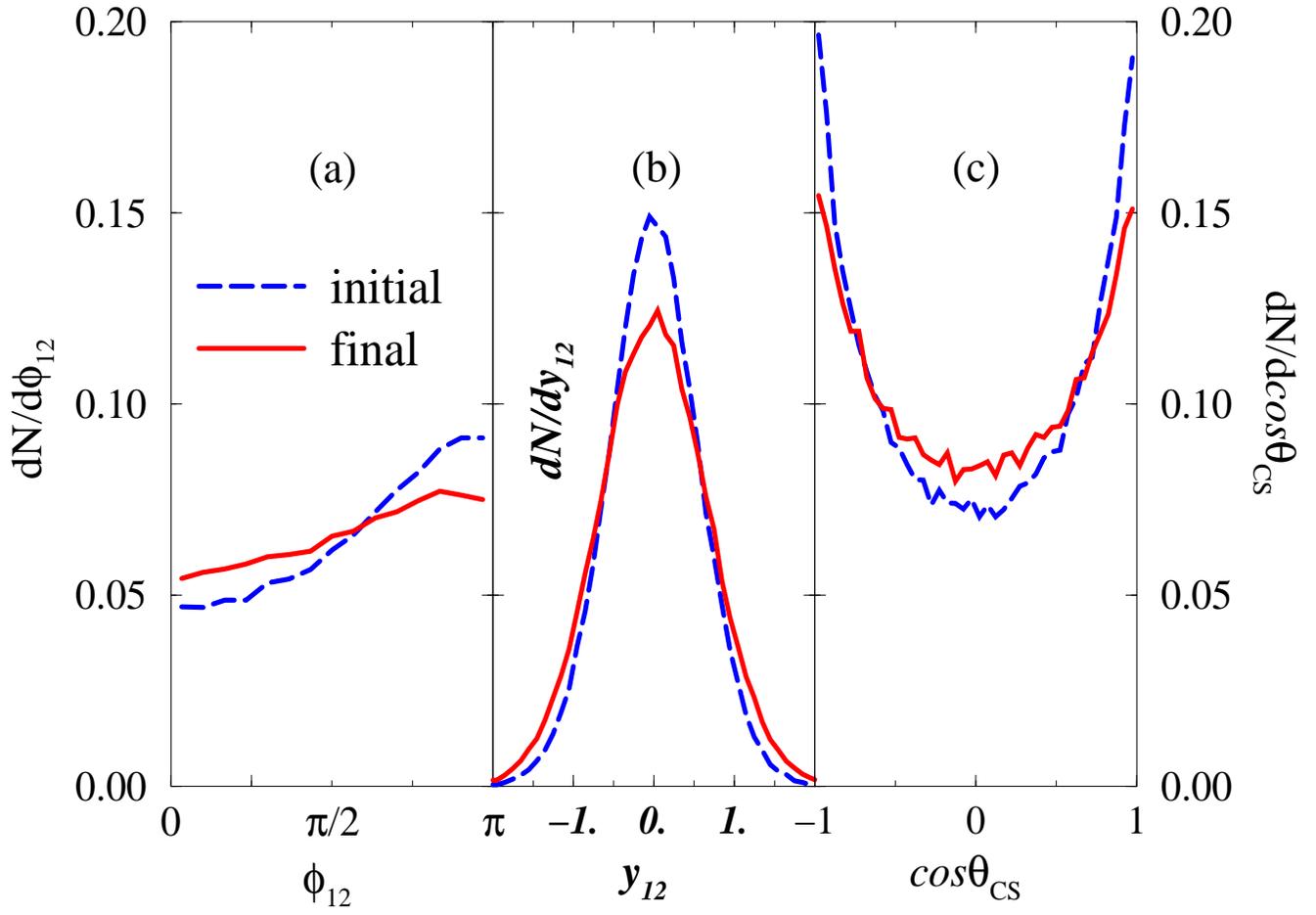}}
\vspace{1cm}
\caption{
Spectra (a) as a function of $\phi_{12}$, the difference between the two
azimuthal angles; (b) as a function of $y_{12}$, the difference between
the two rapidities; and (c) as a function of $\cos \theta_{CS}$, of a
charmed meson pair. 
}
\label{fig2_corr}
\end{figure}

Shown in Fig.\ref{fig1_mty} are the $m_T$ spectra within $-0.5 < y_D < 0.5$ and
the rapidity distribution for initial and final D-mesons. All figures are 
normalized to one central $Pb+Pb$ event at $E_{Lab}=158$ AGeV. As expected,
initial and final D-mesons have roughly the same rapidity spectra, and the
inverse slope of the $m_T$ spectra at small values of ($m_T-m_D$) increases
significantly after the inclusion of the simulated final state rescatterings.
The inverse slope parameter, $T_{eff}$, corresponding to a fit in the region
($m_T-m_D) \in (0,1)$ GeV/c$^2$ increases from $160$ MeV to $290$ MeV. 
To show the correlation of charm and anti-charm particles in a
pair, we plot the azimuthal angle difference, the rapidity difference between
the two and the correlation in $\cos\theta_{CS}$ in Fig.\ref{fig2_corr}, where
$\cos\theta_{CS}$ is the angle of one D meson relative to the beam axis in the
rest frame of the charm pair. 
The correlation in azimuthal angles is reduced by random scatterings, however,
the rapidity gap between the two particles in a charmed meson 
pair remains almost the same.  The distribution in $\cos\theta_{CS}$ would be
flat if charm mesons would follow a static-fireball distribution so that there
would be no correlation between two charm mesons in a pair.
However, the charmed meson pair are still strongly correlated in rapidity with
a big rapidity gap in our model due to the simulated longitudinal flow. 
This is the reason why there is still a strong correlation in
$\cos\theta_{CS}$ despite of final state rescatterings.  
In fact, the correlation in $\cos\theta_{CS}$ still remains strong even if
one removes the initial correlation in azimuthal angles, $\phi_{12}$, of the
charmed meson pair.

For semileptonic decays of D-mesons to muons, the average branching ratio
of $c \rightarrow \mu^+ X$ is taken to be 12\%.  The muon energy spectrum from
$D$-meson decays is consistent with the measurement of the MARK-III
experiment\cite{mark3}.   
Since the number of charm pairs, $N^{PbPb}_{c\bar c}$, is much less than 1, 
the dilepton spectrum from decays of uncorrelated charm pairs is negligible.
In addition, we note that like-sign subtraction is applied in the NA50 data
analysis. 

The dimuon invariant-mass, pair rapidity, and $\theta^{\mu}_{CS}$, which is the
angle of one muon relative to the beam axis in the dimuon rest
frame(Collins-Soper reference frame), are defined as: 
\begin{eqnarray}
M_{\mu^+\mu^-}&=&\sqrt {(p_\alpha^{\mu^+}+p_\alpha^{\mu^-})^2} \nonumber \\
Y_{\mu^+\mu^-}&=&\tanh ^{-1} \beta_{z,pair} \nonumber \\
\cos \theta^{\mu^+}_{CS}&=&\frac{p_z^{\mu^+}-\beta_{z,pair}\;E^{\mu^+}}
{\sqrt{1-\beta_{z,pair}^2} \sqrt{M_{\mu^+\mu^-}^2/4-m_\mu^2}} \;\;. \nonumber
\end{eqnarray}
where $\beta_{z,pair}=(p_z^{\mu^+} +p_z^{\mu^-})/(E^{\mu^+}+E^{\mu^-})$.

\begin{figure}[h]
\setlength{\epsfxsize=\textwidth}
\setlength{\epsfysize=0.6\textheight}
\centerline{\epsffile{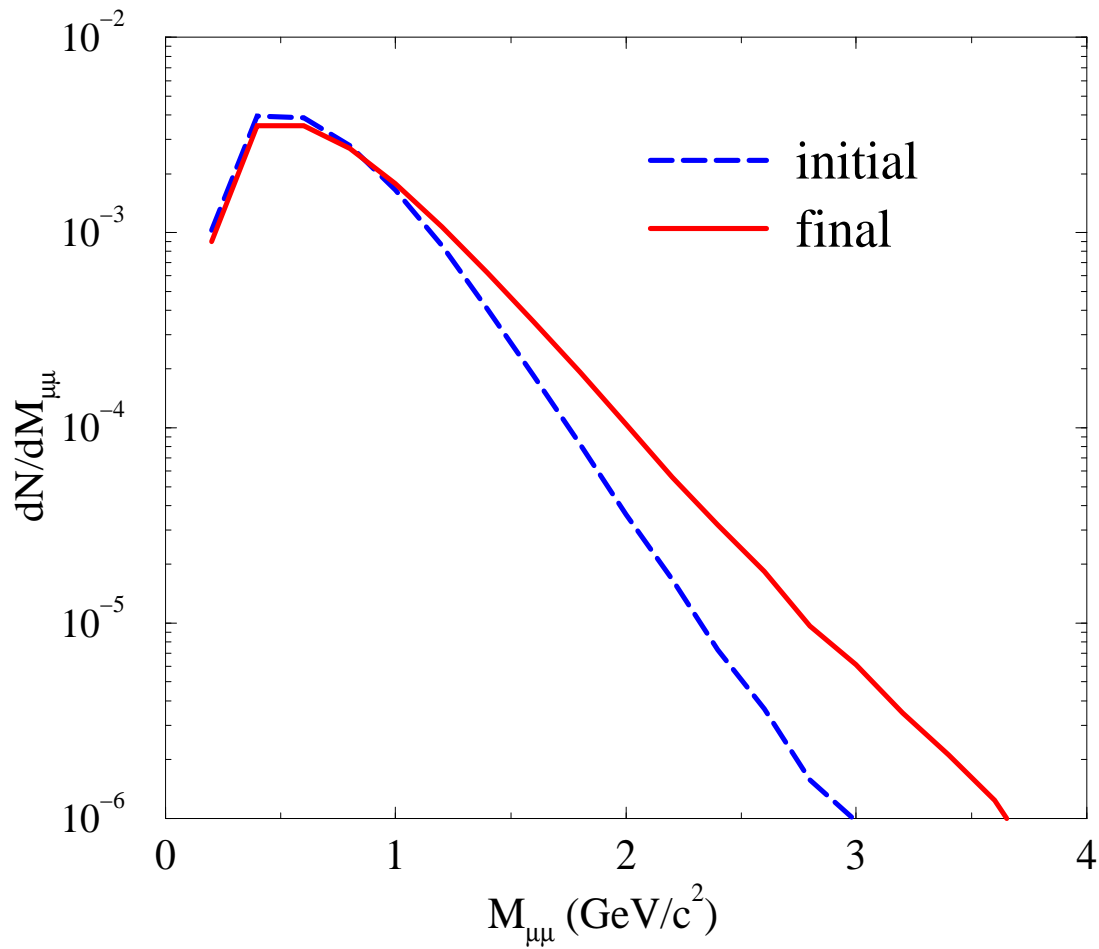}}
\vspace{1cm}
\caption{
Invariant-mass spectra of dimuons from charm pair decays.  The dashed and
solid curves come from D-meson decays without and with final state
rescatterings, respectively.
}
\label{fig3_mass}
\end{figure}

Invariant-mass spectra of dimuons from charm pair decays are
plotted in Fig.\ref{fig3_mass}. The total dimuon yield does not change since
the total charm yield does not change.  However, there are much more high
invariant-mass dimuons after final state rescatterings.  The dimuon pair
rapidity distribution has little change.   

To consider the effect of the limited acceptance of an experiment, we
approximate the NA50 acceptance by taking the following cuts on dimuon
variables:  
\begin{eqnarray}
M_{\mu^+\mu^-} &>& 1.5{\rm \;GeV/c}^2 \;,
Y_{\mu^+\mu^-}^{CMS} \in (0., 1.) \;,
\cos \theta^{\mu^+}_{CS} \in (-0.5, 0.5)\; , \nonumber
\end{eqnarray} 
and cuts on single muons:
\begin{tabbing}
\=$\theta^{\mu}$ \=\ $\in (0.037, 0.108)$, \\
\> $E^{\mu}$ \> $> 8.+1600\times(\theta^{\mu}-0.065)^2$,  when  
$\theta^{\mu}$ \=$ \in (0.037, 0.065)$,\\
\> \> $>8. $,  \> $\in (0.065, 0.090)$,\\
\> \> $> 8.+1300\times(\theta^{\mu}-0.090)^2$, \> $\in (0.090, 0.108)$,
\end{tabbing} 
where $Y_{\mu^+\mu^-}^{CMS}=Y_{\mu^+\mu^-}-2.9$ for $E_{Lab}=158$ AGeV.   

\begin{figure}[h]
\setlength{\epsfxsize=\textwidth}
\setlength{\epsfysize=0.6\textheight}
\centerline{\epsffile{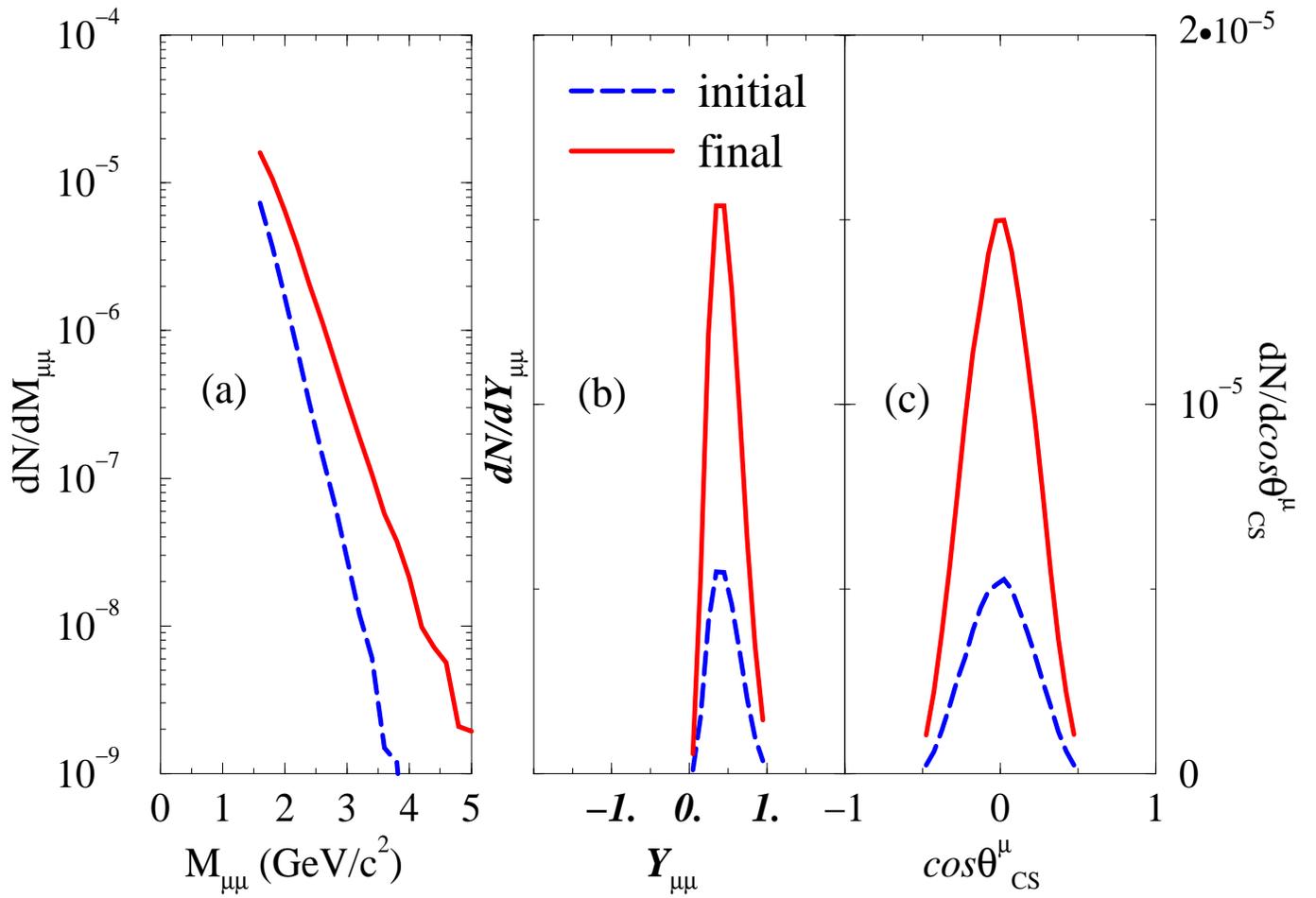}}
\vspace{1cm}
\caption{
(a) Invariant-mass, (b) pair rapidity and (c) $\cos \theta^{\mu^+}_{CS}$
spectra of dimuons from charm pair decays inside the approximate NA50
acceptance. 
}
\label{fig4_na50}
\end{figure}

The invariant-mass, pair rapidity and $\cos \theta^{\mu^+}_{CS}$ spectra inside
this NA50 acceptance are shown in Fig.~\ref{fig4_na50}.
Due to the broadening of the D-meson $m_T$ spectrum, the dimuon yield from
D-meson decays in the above acceptance is enhanced significantly.  The
enhancement is stronger at larger dimuon invariant masses, reflecting the
relative change of the $m_T$ spectrum. 
However, the shapes of the pair rapidity 
spectrum and the $\cos \theta^{\mu^+}_{CS}$ spectrum stay roughly the same.  
We define the enhancement factor of the dimuon yield from open charm decays
inside the acceptance, $R$, as the ratio between accepted dimuon yields after
and before final state rescatterings. It is equivalent to the area under the
solid curves divided by the area under the dashed curves in
Fig.~\ref{fig4_na50}-b and c. With a value of $T=150$ MeV for final state
rescatterings which corresponds to a D-meson $m_T$ inverse slope of $290$ MeV,
one obtains an enhancement factor about $R=3.0$. 

\begin{figure}[h]
\setlength{\epsfxsize=\textwidth}
\setlength{\epsfysize=0.6\textheight}
\centerline{\epsffile{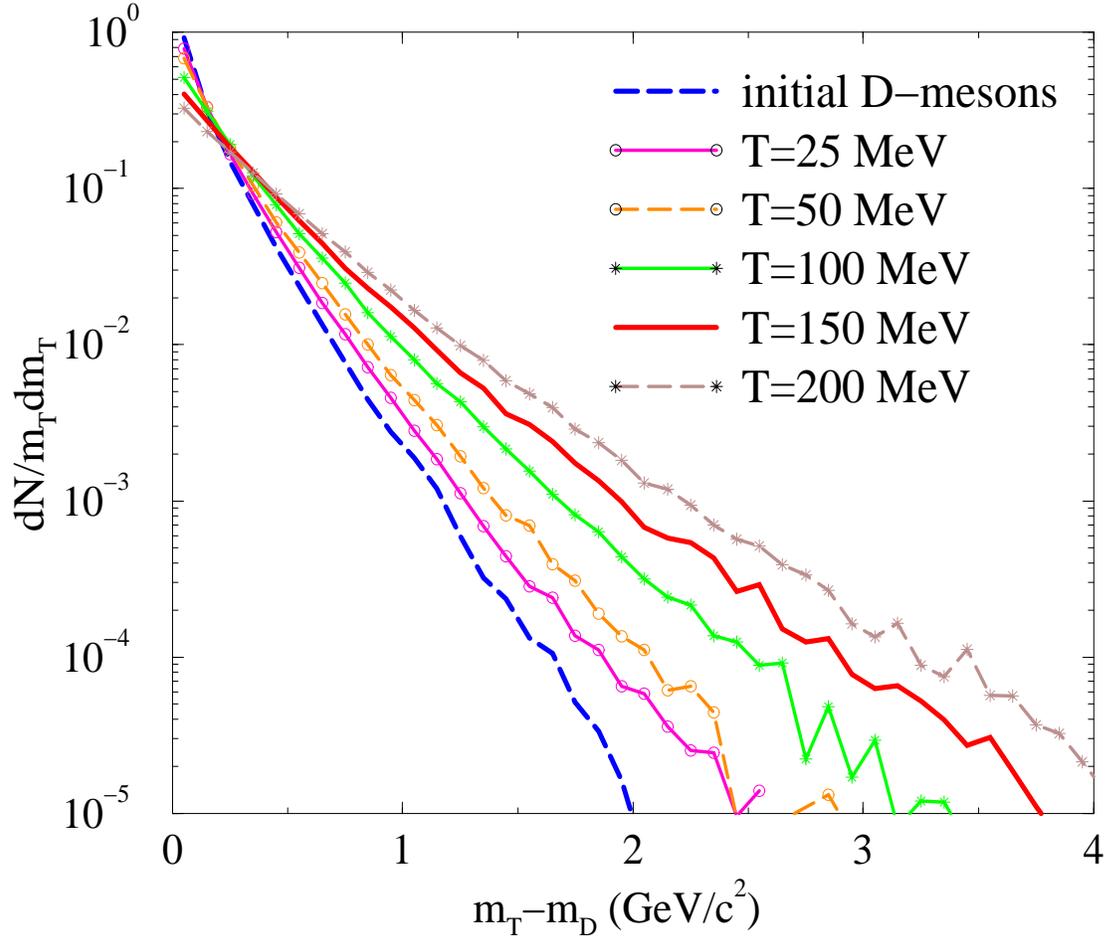}}
\vspace{1cm}
\caption{
$m_T$ spectra of D-mesons at mid-rapidity 
after different final state rescatterings.  
}
\label{fig5_mt}
\end{figure}
\begin{figure}[h]
\setlength{\epsfxsize=\textwidth}
\setlength{\epsfysize=0.6\textheight}
\centerline{\epsffile{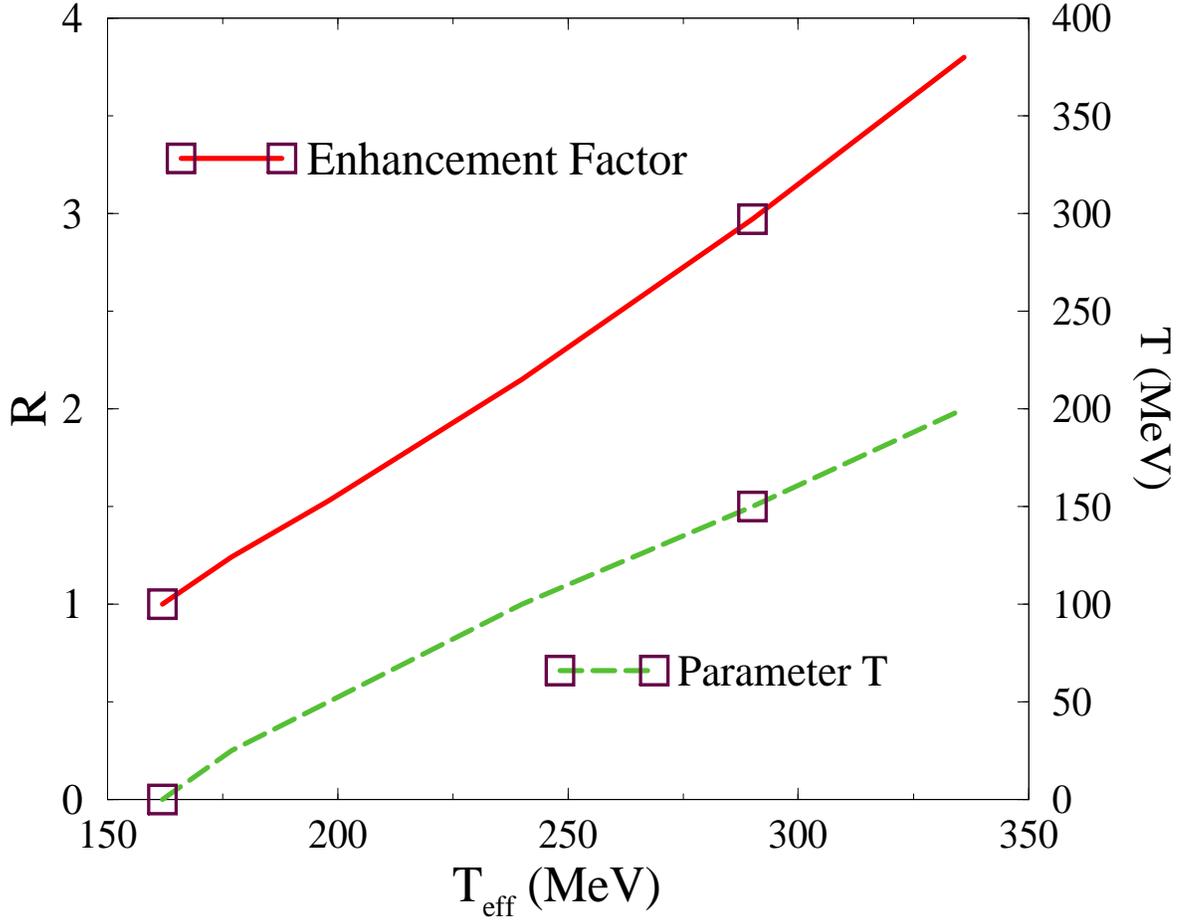}}
\vspace{1cm}
\caption{
The dimuon enhancement factor, $R$, and the parameter $T$, as a function of 
the D-meson inverse slope, $T_{eff}$.  The two boxes on the left refer to the
case without final state rescatterings, while the two boxes on the right refer
to the default case of final state rescatterings with $T=150$ MeV.
}
\label{fig6_r}
\end{figure}

As we have demonstrated so far, the enhancement factor of IMR dimuons from
charmed meson decays, $R$, depends on the D-meson transverse 
momentum spectrum characterized by $T_{eff}$.  This inverse slope is 
related to the parameter $T$, which in our model controls the strength of
final state rescatterings.  As one can easily imagine that the degree of
final state rescatterings must depend on the size of the dense matter in
heavy-ion collisions which in turn depends on the impact parameter. In
principle one can study the effect of rescatterings in a cascade model and
study the dependence of the effective $T_{eff}$ or the enhancement factor $R$
on the centrality of heavy-ion collisions. To demonstrate such possible
dependence, we simply vary the effective local temperature $T$ here and study
how the effective IMR dilepton spectra will change. Show in Fig.\ref{fig5_mt}
are the final D-meson $m_T$ spectra for different values of temperature $T$.
The inverse slopes, $T_{eff}$, from these spectra can 
also be read off in Fig.~\ref{fig6_r}, where the relations $R(T_{eff})$ and
$T(T_{eff})$ are shown.  
The two boxes on the left in Fig.~\ref{fig6_r} refer to the expected dimuon
yield by simply scaling up $pp$ results without consideration of final state
rescatterings, thus $T=0$ and $R=1$ by definition.  
The two boxes on the right refer to the enhanced dimuon yield due to
final state rescatterings with a default value of $T=150$ MeV.  
Both the $R(T_{eff})$ and $T(T_{eff})$ relation are roughly
linear, thus the $R(T)$ relation is also almost linear. However, the relation
between $T$ or $T_{eff}$ and the impact parameter could be more complex, 
which we will not address in this paper. 

We would like to point out that the absolute value of the total charm cross
section, $\sigma^{pp}_{c\bar c}$, 
is not important for our study of the enhancement factor, $R$, for dimuons from
charmed meson decays; 
because we normalize $R$ to $1$ for $pA$ collisions and thus
$\sigma^{pp}_{c\bar c}$ cancels.  We take $\sigma^{pp}_{c\bar c}$ to be
$6.8\mu$b just to be consistent with the NA38 data\cite{carlos}.

Besides final state rescatterings, there is also some effect from initial
multiple scatterings, which effectively increases the parton intrinsic $k_t$.  
By modeling initial multiple scatterings in a random-walk picture where the
additional average $k_t^2$ is proportional to the system size, 
we checked that this effect gives a dimuon enhancement less than $10\%$.  
Anti-shadowing could be a source for the enhancement too, because at SPS
energies the charm production probes the region with Bjorken $x \sim 0.2$,
where the gluon anti-shadowing could be close to its maximum
value\cite{eskola}. However, this effect is also checked to be no more than
$20\%$. 
  

In summary, the effect of final state rescatterings on D-mesons in $Pb+Pb$
collisions at SPS energies are studied. We found that the dimuon yield from
open charm decays inside the NA50 acceptance is significantly enhanced as a
result.
Our model for the effects of final state rescatterings is very schematic.
However, the essential point is the broadening of D-meson $m_T$ spectra, which
can be roughly represented by the increase of $T_{eff}$, the inverse slope of
the D-meson $m_T$ spectrum. More studies can be done beyond this schematic
model.  
The centrality dependence of the enhancement can be studied and compared to
experimental data at various values of total $E_T$.
The $p_\perp$ dependence of the enhancement can also be studied by applying
various $p_\perp$ cuts, and a larger enhancement is expected when more
stringent energy cut is applied to single muons. However, the 
ultimate test of this model should come from direct measurements of D-meson
spectra. 

Acknowledgments: We are grateful to the Institute for Nuclear Theory at the
                 University of Washington for supporting the INT-98-1 program
                 where this work was started.  
                 Special thanks are due to C. Gale for numerous valuable
                 discussions  during the program.  
                 We also thank C. Louren\c{c}o, K. Redlich, H. Satz,
                 E. Scomparin and N. Xu for helpful discussions.   

\pagebreak

\pagebreak

\pagebreak

\pagebreak

\pagebreak

\pagebreak

\pagebreak
{}

\end{document}